\documentclass[9pt,twocolumn,twoside]{opticajnl}
\journal{opticajournal} 

\setboolean{shortarticle}{true}

\usepackage{lineno}
\usepackage{upgreek}

\title{Dual-color Q-switched mode-locking in an Erbium-doped fiber laser}

\author[1,2]{Chenyue Lv}
\author[1,2,$\dagger$]{Baole Lu}
\author[1,2,*]{Jintao Bai}

\affil[1]{State Key Laboratory of Energy Photon-Technology in Western China, International Collaborative Center on Photoelectric Technology and Nano Functional materials, Institute of Photonics \& Photon-technology, Northwest University, Xi’an 710127, China}
\affil[2]{Shaanxi Engineering Technology Research Center for Solid State Lasers and Application, Provincial Key Laboratory of Photo-electronic Technology, Northwest University, Xi’an 710127, China}

\affil[$\dagger$]{lubaole1123@163.com}
\affil[*]{baijt@nwu.edu.cn}

\begin{abstract}
Q-switched mode-locking (QML) has been widely observed in various lasers, but its generation mechanism in passive mode-locking remains unclear. In this paper, we build up a dual-color QML Erbium-doped fiber laser and find a bound-state-like envelope on the optical spectrum for the first time. Theoretically, the formation mechanism of QML is numerically investigated using the coupled Ginzburg-Landau equations. In addition, we demonstrated the existence of two QML pulse evolution patterns with gain or polarization state variations in simulation. Our results deepen the understanding of QML pulses in mode-locked fiber lasers and provide a foundation for studying mode-locking nonlinear evolutionary paths. \href{https://opg.optica.org/content/author/portal/item/review-copyright-permissions/cpyrt-lic-statements}{copyright licensing statement}.
\end{abstract}

\setboolean{displaycopyright}{false} 

\begin{document}

\maketitle

Passive mode-locked fiber lasers are widely used in micromachining \cite{1}, biomedicine \cite{2}, optical communications \cite{3}, etc., which is one of the frontier hot research directions in the ultrafast laser area. The pulses in a mode-locked fiber laser are known as optical solitons arising from the cavity's combined balance among dispersion, nonlinearity, gain, and loss \cite{4}. The emergence of solitons is accompanied by many complex nonlinear dynamics \cite{5,6,7,8,9,10,11}. With the continuous improvement of detection technology, especially real-time measurement, those nonlinear phenomena have been transiently characterized \cite{12,13,14}. Also, the buildup processes of various solitons are clarified by the time-stretching dispersive Fourier transform \cite{15,16}. Even with continuous investigation, many dynamic processes of soliton nonlinear phenomena remain unclear. For example, the soliton buildup process in the case of near-zero dispersion and the Q-switched mode-locking (QML). In the QML state, the peak power of the pulse obtains more than the multiple of a continuous wave mode-locking (CWML) \cite{17,18,19,20}. However, most studies have focused on actively QML \cite{21,22}, and there is no satisfactory model describing passively QML so far.

In this study, we investigated a dual-color QML Erbium (Er)-doped fiber laser. Bound-state-like envelopes were found on the QML optical spectrum for the first time. Theoretically, we built a set of the coupled Ginzburg Landau equations (CGLEs) based on the hybrid mode-locked fiber laser and numerically solved them using the fourth-order Runge-Kutta (RK4) method to obtain the evolution of the QML pulses. The simulation results verify the periodic oscillations of the QML pulse spectrum, and two evolution patterns of the QML pulses are found while changing the gain and polarization states. In addition, we numerically simulate the bound-state-like modulation spectrum by modeling the sampling process of the spectrometer. Our results deepen the understanding of the QML mode-locking mechanism and provide a foundation for studying nonlinear evolutionary paths in fiber lasers.

The dual-color QML mode-locked Er-doped fiber laser shown in Fig. 1 was built to generate dual-color Q-switched mode-locked solitons based on saturable absorber (SA) and nonlinear polarization rotation (NPR). The mode-locked fiber laser consists of a 2.00 m Er-doped fiber (EDF, Er110-4/125, LIEKKI), a 1.85 m dispersion compensating fiber (DCF, DCF38), and two sections of polarization maintaining fiber (PMF, PM1550-XP). Two polarization controllers (PCs) are placed in the fiber laser to adjust the polarization state accurately. In addition, a polarization-dependent isolator (PD-ISO) ensures unidirectional laser transmission within the cavity; a 10:90 optical coupler (OC) is used with a  90$\%$ port connected into the cavity and a 10$\%$ port for pulse output. The total length of the cavity is 10.39 m, corresponding to a fundamental rate of 19.64 MHz. The group velocity dispersions are estimated to be $\sim$ 11.4 $\rm ps^{2}/km$ and $\sim$ 62.5 $\rm ps^{2}/km$ for EDF and DCF, respectively, and $\sim$ -22.8 $\rm ps^{2}/km$ for single-mode fiber (SMF, SMF-28e+). The net cavity dispersion was calculated as -0.0105 $\rm ps^{2}$, a slightly negative of the near-zero dispersion. For output pulse measurement, a high-speed real-time oscilloscope (Agilent, DSO9104A) is used to track temporal information, an optical spectrum analyzer (Yokogawa, AQ6370C) to record time-averaged optical spectrum, and a radio frequency (RF) spectrum analyzer (Keysight, N9000B CXA signal analyzer) for RF spectrum.

\begin{figure}[ht]
	\centering
	\fbox{\includegraphics[width=\linewidth]{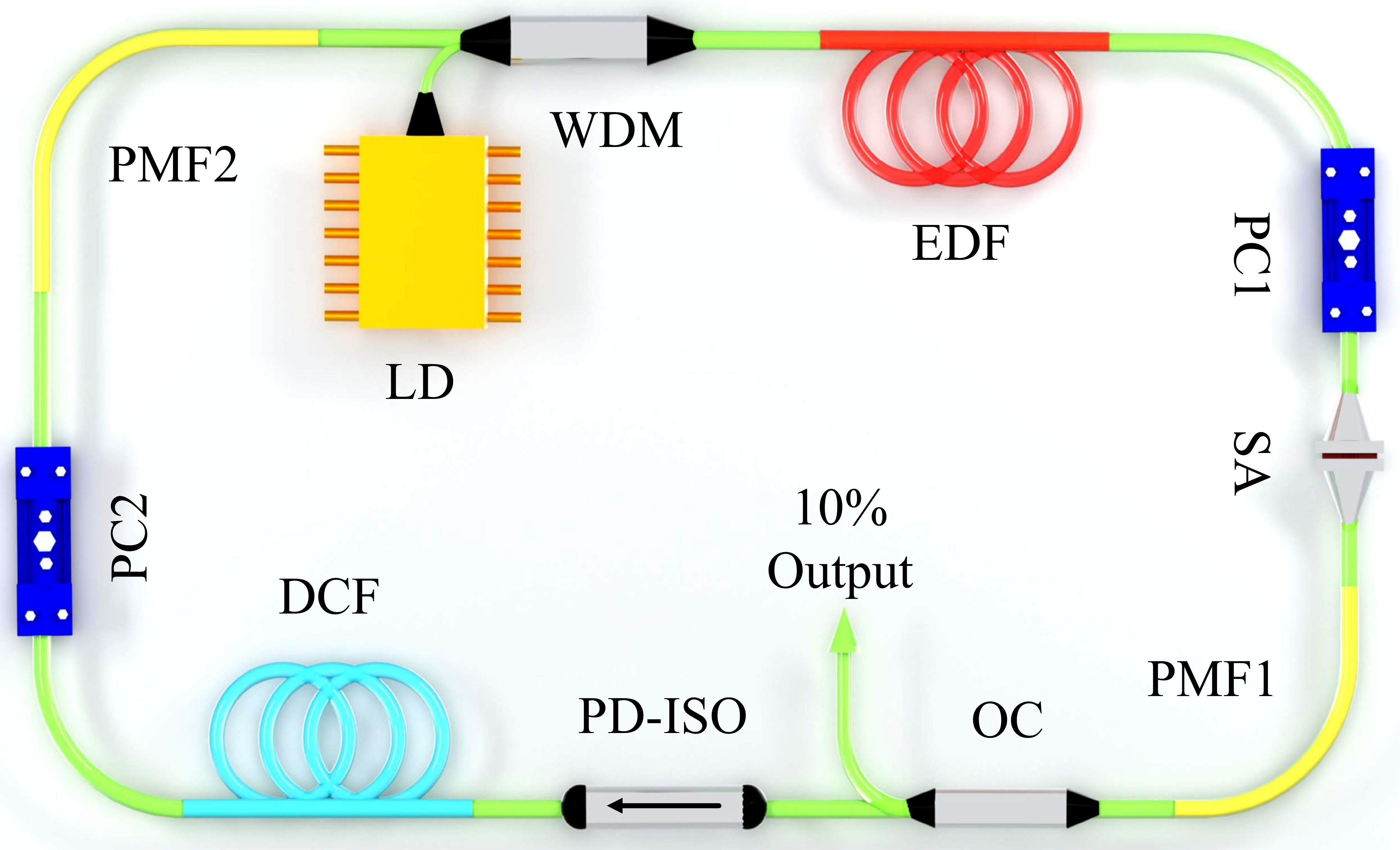}}
	\caption{Schematic of near-zero net cavity dispersion dual-color QML Er-doped fiber laser.}
\end{figure}

The mode-locked threshold was approximately 200 mA for the laser shown in Fig. 1. It should be noted that this threshold may change at a range of ~20 mA when the SA is re-inserted into the laser. Increasing the pump power to 300 mA, dual-color QML pulses can be observed in the hybrid mode-locked fiber laser, as shown in Fig. 2. Figures. 2(a)-(c) present progressively detailed pulse trains. A 50 $\upmu$s QML train of three Q-envelopes is shown in Fig. 2(a), while Fig. 2(b) illustrates a zoomed-in single Q-envelope with 2.19 $\upmu$s. Figure 2(c) shows the mode-locked pulse train in the Q envelope, while Fig. 2(d) presents a single mode-locked pulse. The QML pulse width is calculated from Fig. 2(d) of 200 ps since it is difficult to capture the autocorrelation trace due to the large variation in the intensity of neighboring pulses in the QML. 

From the optical spectrum in Fig. 2(e), it can be found that the QML pulse output dual wavelength is at 1563.00 nm and 1571.08 nm and the wavelength difference is 8.08 nm. A 0.56 nm bound-state-like modulation envelope exists at both wavelengths, and there is no significant superimposed interference in the overlapping part of the dual-wavelength envelope. More impressively, this bound-state-like modulation shows an alternating behavior over time when the scanned interval is at 1 second, as shown in Fig. 2(g). To clarify the dynamic behavior, the optical spectrums at 72 s and 73 s are given in Fig. 2(h), and the detail diagrams for the two wavelengths are displayed separately. In the zoomed-in diagrams, the two roundtrips of the optical spectrum just 1 s apart in scan time produce a variation of about half a modulation period, which is sufficient evidence for the QML pulse's oscillatory behavior. The radio frequency (RF) spectrum of the QML pulse is demonstrated in Fig. 2(f). The repetition frequency of the Q-envelope is clarified as 50 kHz over a swept range of 1 MHz. The inset shows the corresponding mode-locked RF spectrum at a fundamental rate of 19.635 MHz with a signal-to-noise ratio of 50.37 dB, similar to the theoretical calculation. In addition, the fundamental rate in the measurement range of 1.6 MHz is a typical QML spectrum with multiple satellite peaks at intervals of 50 kHz.

\begin{figure}[ht]
	\centering
	\fbox{\includegraphics[width=\linewidth]{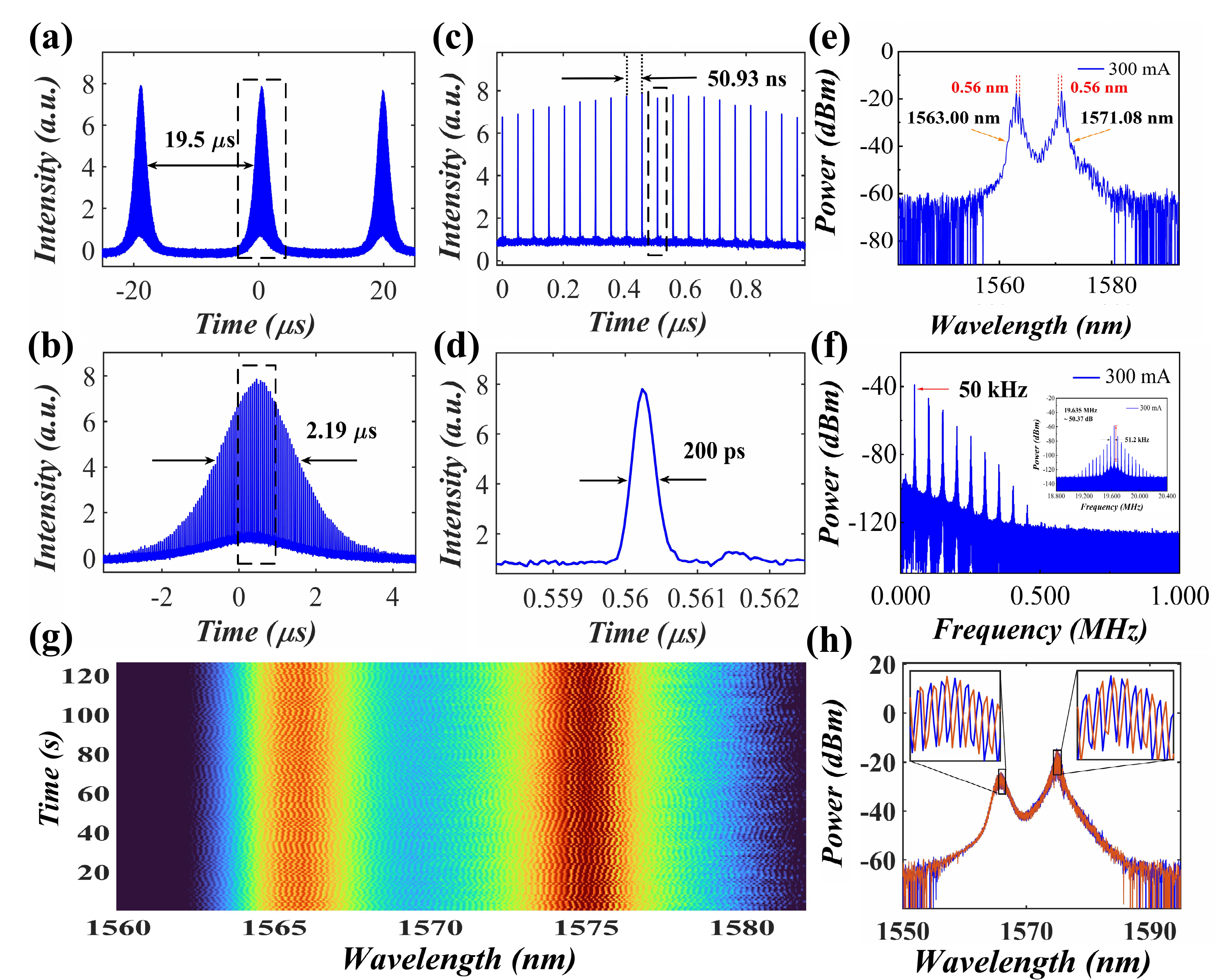}}
	\caption{Dual-color QML operation. (a)-(c) typical QML pulse train and (d) single QML pulse; (e) optical spectrum; (f) the RF spectrum; (g) 1-second intervals optical spectrums; (h) optical spectrums at 72 s and 73 s.}
\end{figure}

Increase the pump power from 255 mA at 1 mA intervals to 310 mA. As shown in Fig. 3(a), the bound-state-like modulation envelope interval on the optical spectrum decreases from 1.12 nm to 0.16 nm, while the dual center wavelength of the optical spectrum has almost no change. The corresponding RF spectrum is exhibited in Fig. 3(b), from which the repetition rate of the Q-switched envelope gradually increases. The trend of the RF spectrum with power is consistent with the general Q-switched pulse and opposite to the envelope effect on the optical spectrum. The bound-state-like modulation envelopes on the optical spectrum are instead closer when the Q-switched envelopes are approaching each other. And this is proof that modulation intervals on the optical spectrum are not due to the Q-switched.

In addition, the envelope intervals can also vary with polarization state change as Fig. 3(c) shows. With the continuous change of the polarization state, the intervals of the bound-state-like modulation firstly decrease from 0.68 nm to 0.16 nm then back to 0.28 nm, and the center wavelength position is changed. Due to the manual squeezing PC used in the experiment, it is unavoidable to generate a shift in the intracavity loss when tuning the polarization state. This loss change causes the modulation interval to rise and the two center wavelengths of the QML move away from each other. Based on the above experimental results, the bound-state-like modulation envelope on the optical spectrum may be related to the changing rate of the mode-locked pulses in QML. In other words, the faster the mode-locked pulses change in the QML, resulting in the smaller the bound-state-like envelope interval.

\begin{figure}[ht]
	\centering
	\fbox{\includegraphics[width=\linewidth]{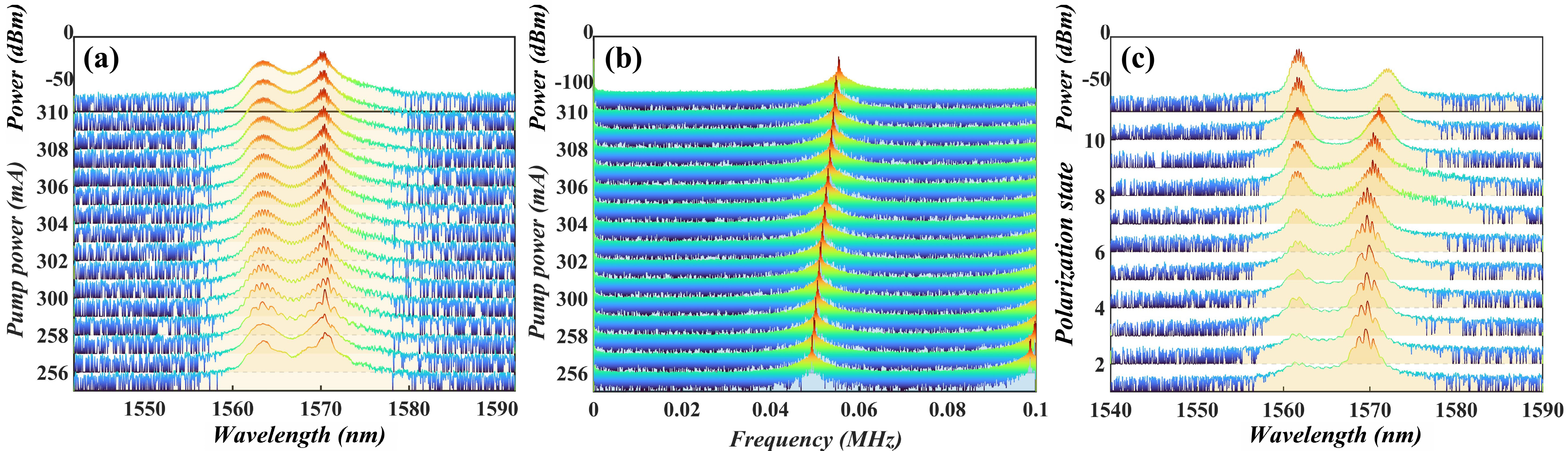}}
	\caption{Evolution of dual-color QML pulse optical spectrums with (a) increasing the pump power and (c) finely tuning the polarization state; (b) corresponding RF spectrum to (a).}
\end{figure}

To further investigate the appearance of bound-state-like modulation envelope in the optical spectrum, we have theoretically constructed a laser model based on the CGLEs and performed numerical simulations. The equations are shown below:

\begin{equation}
\begin{aligned}
	\frac{\partial \varphi_{x,y}}{\partial z} =  -i \frac{\beta_{2}}{2} \frac{\partial^{2} \varphi_{x,y}}{\partial t^{2}} + i\gamma \left(\left|\varphi_{x,y}\right|^{2} +  \frac{2}{3}\left|\varphi_{y,x}\right|^{2} \right) &  \varphi_{x,y}\\
	+\frac{i \gamma}{3} \varphi_{x,y}^{*} \varphi_{y,x}^{2}+\frac{g-l}{2} \varphi_{x,y}+\frac{g}{2 \Omega_{g}^{2}} \frac{\partial^{2} \varphi_{x,y}}{\partial t^{2}}\\
\end{aligned}
\end{equation}

In addition, a Gaussian-type complex bandpass filter with two center wavelengths is used to simulate the composite filter model \cite{23}. Using transmittance function and transmission matrix to simulate the SA effect, the NPR effect, and other intracavity elements to build up the fiber laser. The model remains essentially the same as in reference \cite{23}. The only difference is the angle between the main optical axes of the two PMFs is additionally taken into account as follows \cite{6}:

\begin{small} 
\begin{equation}
	\left[\begin{array}{l}
		\varphi_{outx } \\
		\varphi_{outy }
	\end{array}\right]=J_{\theta}
    \left[\begin{array}{l}
		\varphi_{inx} \\
		\varphi_{iny} 
	\end{array}\right]=
    \left[\begin{array}{cc}
	      \cos (\alpha) & -\sin (\alpha) \\
	      \sin (\alpha) & \cos (\alpha)
    \end{array}\right]
    \left[\begin{array}{l}
          \varphi_{inx} \\
          \varphi_{iny} 
    \end{array}\right]
\end{equation}
\end{small}

The parameters used in the numerical simulations correspond to the fiber laser in the experiment, shown in Table 1.

\begin{table}[htbp]
	\centering
	\caption{\bf Parameters in the numerical simulation of dual-color Q-switched 
		mode-locking Er-doped fiber laser.}
  	\resizebox{8.8cm}{!}{
	\begin{tabular}{cccc}
	\hline
	Variable name &Value & Variable name & Value \\
	\hline
	$\beta_{2SMF}$ & $-22.8\: ps^{2}km^{-1}$ & $\beta_{2EDF}$ & $11.4\: ps^{2}km^{-1}$  \\
    $\beta_{2PMF}$ & $-22.0\: ps^{2}km^{-1}$ & $\beta_{2DCF}$ & $62.5\: ps^{2}km^{-1}$  \\
	$\gamma_{PMF}$ & $1.3 \: W^{-1}km^{-1}$ & $\gamma_{EDF}$ & $4.7\: W^{-1}km^{-1}$ \\
	$g_{0}$ & $Varying$ & $P_{sat}$ & $30\: W$\\
	$l$ & $0.15\: m^{-1}$ & $\Omega_{g}$ & $40\: nm$ \\
	$A_{1}$ & $1$ & $A_{2} $ & $1$ \\
	$\lambda_{1}$ & $1562\: nm$ & $\lambda_{2}$ & $1572\: nm$\\
	$\Delta \lambda_{1}$ & $6\: nm$ & $\Delta \lambda_{2}$ & $10\: nm$ \\
	$\alpha_{PC1}$ & $75^{\circ}$ & $\beta$ & $60^{\circ}$ \\
	$\alpha_{PC2}$ & $35^{\circ}$ & $\theta$ & $Varying$\\
	\hline
    \end{tabular}
    }
\end{table}

The intracavity gain and polarization state are varied by changing the values of g$_{0}$ and $\theta$ to simulate the conditions in the experiment. By solving the CGLEs in Matlab using the RK4 method, we obtain the mode-locking process of the dual-color QML pulse. Firstly, the polarization state was kept at 30$^{\circ}$ and the small signal gain g$_{0}$ was increased from 28.0599 to 28.0602, as shown in Fig. 4. Where the three columns correspond to the pulses evolution train, the corresponding optical spectrums evolution, and the peak power curves. Due to the introduction of dispersion compensation, the fiber laser operates at near-zero dispersion. Combined with the dual-filtering effect of the neighboring center wavelength, the pulse obtained from the numerical simulation shows a synchronous dual-wavelength, which is consistent with the experimental QML pulse without interference peak state.

\begin{figure}[ht]
	\centering
	\fbox{\includegraphics[width=8.3cm]{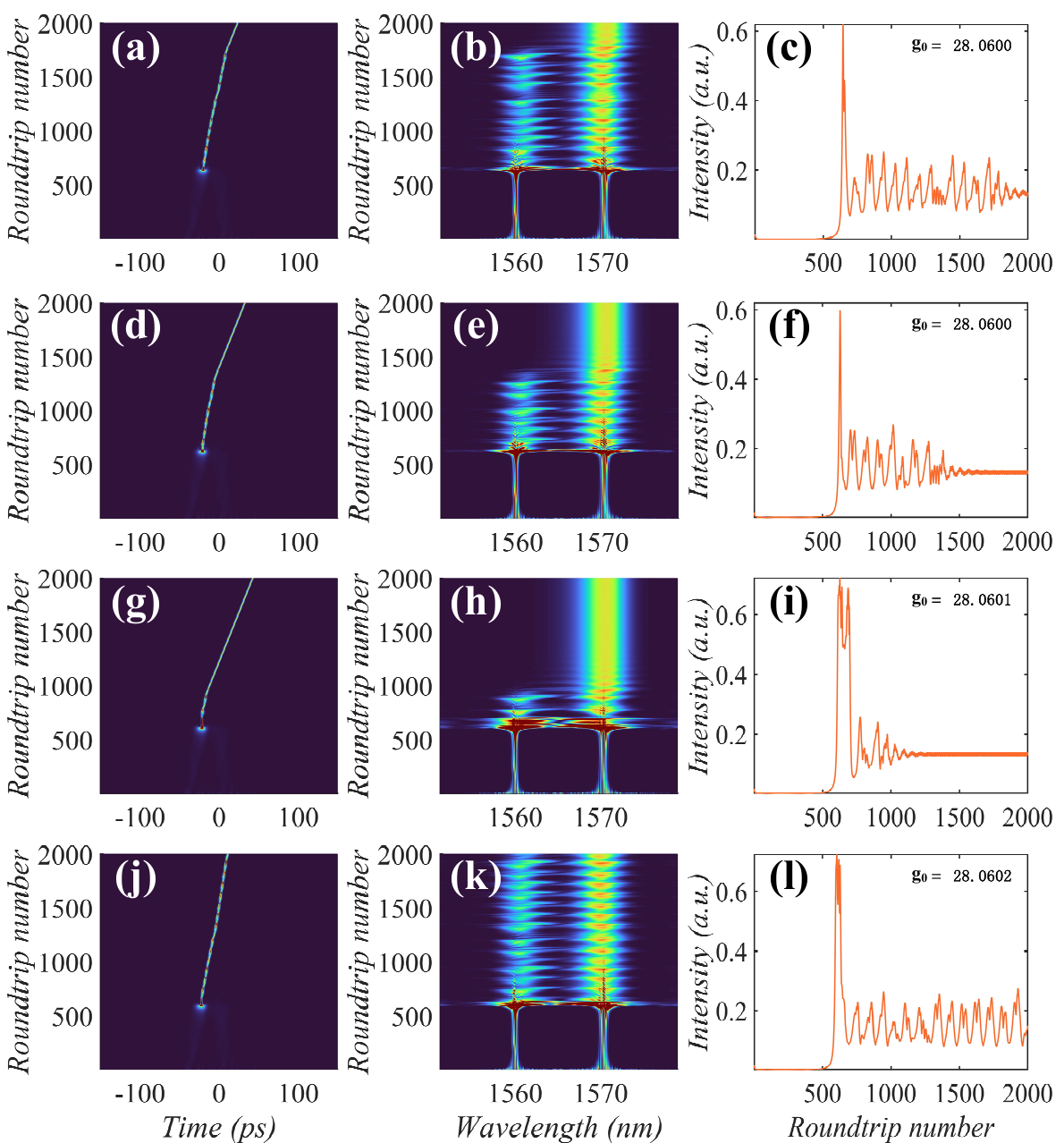}}
	\caption{Numerical simulation results of the dual-color QML fiber laser with different small signal gains. (a)-(c) g$_{0}$ = 28.0599; (d)-(f) g$_{0}$ = 28.0600; (g)-(i) g$_{0}$ = 28.0601; (j)-(l) g$_{0}$ = 28.0602.}
\end{figure}

Take small signal gain as g$_{0}$ = 28.0600 as an example for analysis. From Figs. 4(d)-(f), it can be seen that with the increasing number of roundtrips, both the time domain and the optical spectrum follow up a beating dynamic. Then comes a periodic intensity change, which means that the Q-switched envelope is generated. The subsequent evolution process can be divided into two patterns.  

When the intracavity energy is not enough to generate CWML pulses at both wavelengths at the same time, the QML process can be considered as a struggle process between the two center wavelengths to rob the energy. This struggle continues until one of them wins, and such a QML process will be shorter as the total intracavity energy is greater. In Fig. 4(e), after a period of QML, due to the imbalance of filtering intensity, the pulse with 1572 nm wavelength gradually seizes the cavity energy. That led to the obliteration of the 1562 nm wavelength peak, thus evolving into a stable single-wavelength mode-locked pulse. As the power increases, the time for QML evolving into CWML decreases and the interval of the Q-switched envelope decreases. 

In contrast, the second pattern is that the intracavity energy is sufficient to support dual-wavelength CWML generation, and then the QML competition in the cavity will continue until it becomes a stabilized dual-wavelength CWML. Obviously, this equilibrium between the energies is not so easy to achieve, and the laser is in the QML stage continuously for a long time. By the way, this is also consistent with experimental results in which some unstable QML pulses evolve autonomously into single- or dual-wavelength CWML pulses in a few minutes.

\begin{figure}[ht]
	\centering
	\fbox{\includegraphics[width=8.3cm]{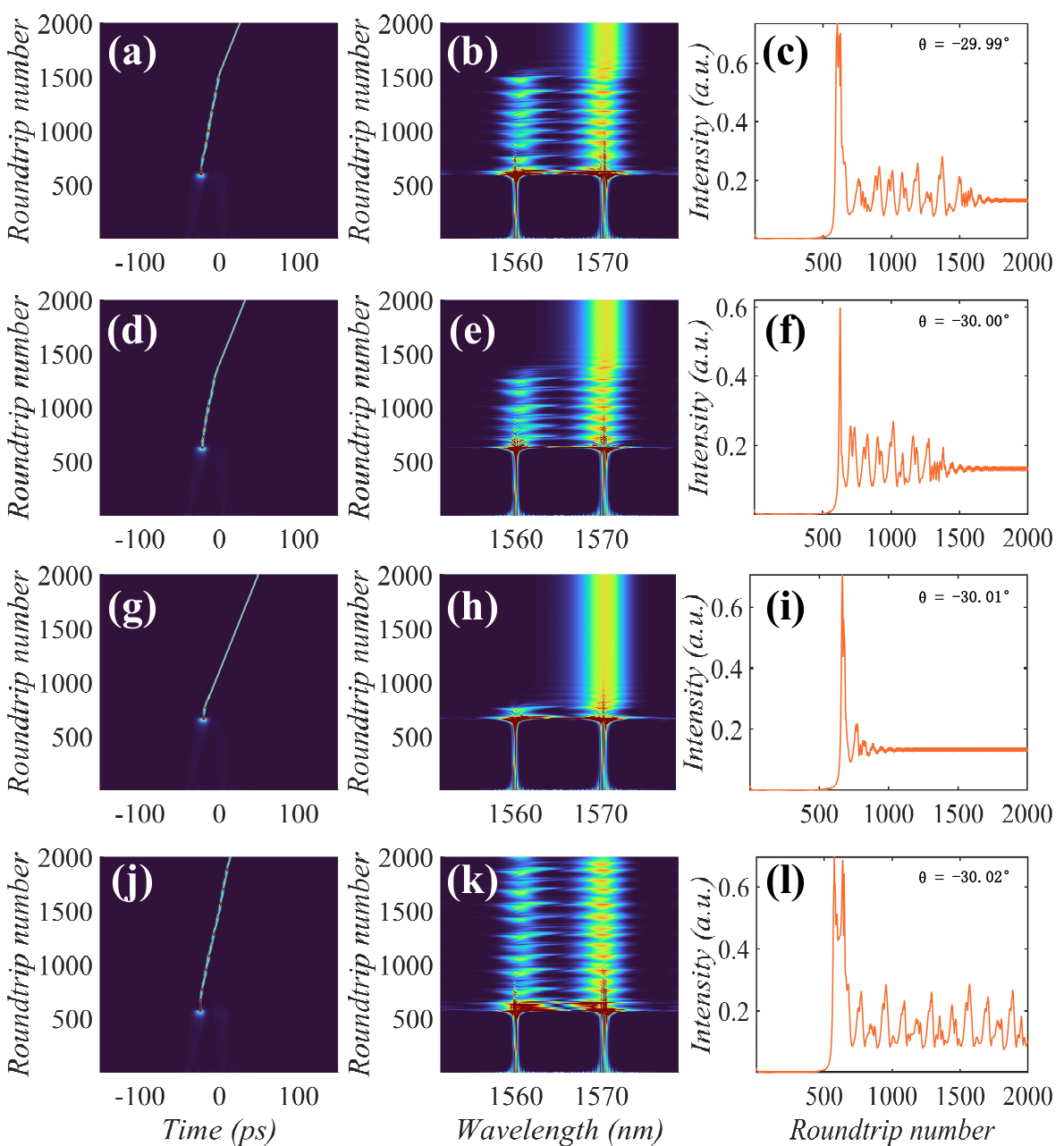}}
	\caption{Numerical simulation results of the dual-color QML fiber laser with different polarization states. (a)-(c) $\theta$ = -29.99$^{\circ}$; (d)-(f) $\theta$ = -30.00$^{\circ}$; (g)-(i) $\theta$ = -30.01$^{\circ}$; (j)-(l) $\theta$ = -30.02$^{\circ}$.}
\end{figure}

In addition, we changed the polarization state of the cavity by varying the magnitude of $\theta$. The numerical simulation results are presented in Fig. 5. Where the negative sign before the $\theta$ value represents that the rotation direction is counter-clockwise. As the $\theta$ value rotates from -29.99$^{\circ}$ to -30.01$^{\circ}$, the QML pulse gradually devolves into the CWML pulse and the evolution time gradually decreases. The whole evolution process can be observed by comparing the vertical columns in Figs. 5(c), 5(f), 5(i), and 5(l). Unlike the experiment, the center wavelength does not change with the polarization state, which might be caused by the inability of the dual-filter model used in the numerical simulation to fully model the squeezed loss with the polarization controller. Combining the simulation results of Fig. 4 and Fig. 5, it can be found that increasing the power and changing the polarization state have almost the same tendencies in the QML pulse evolution process. This identical trend also matches the evolution trend of the bound-state-like envelope in the experimental results.

\begin{figure}[ht]
	\centering
	\fbox{\includegraphics[width=8.3cm]{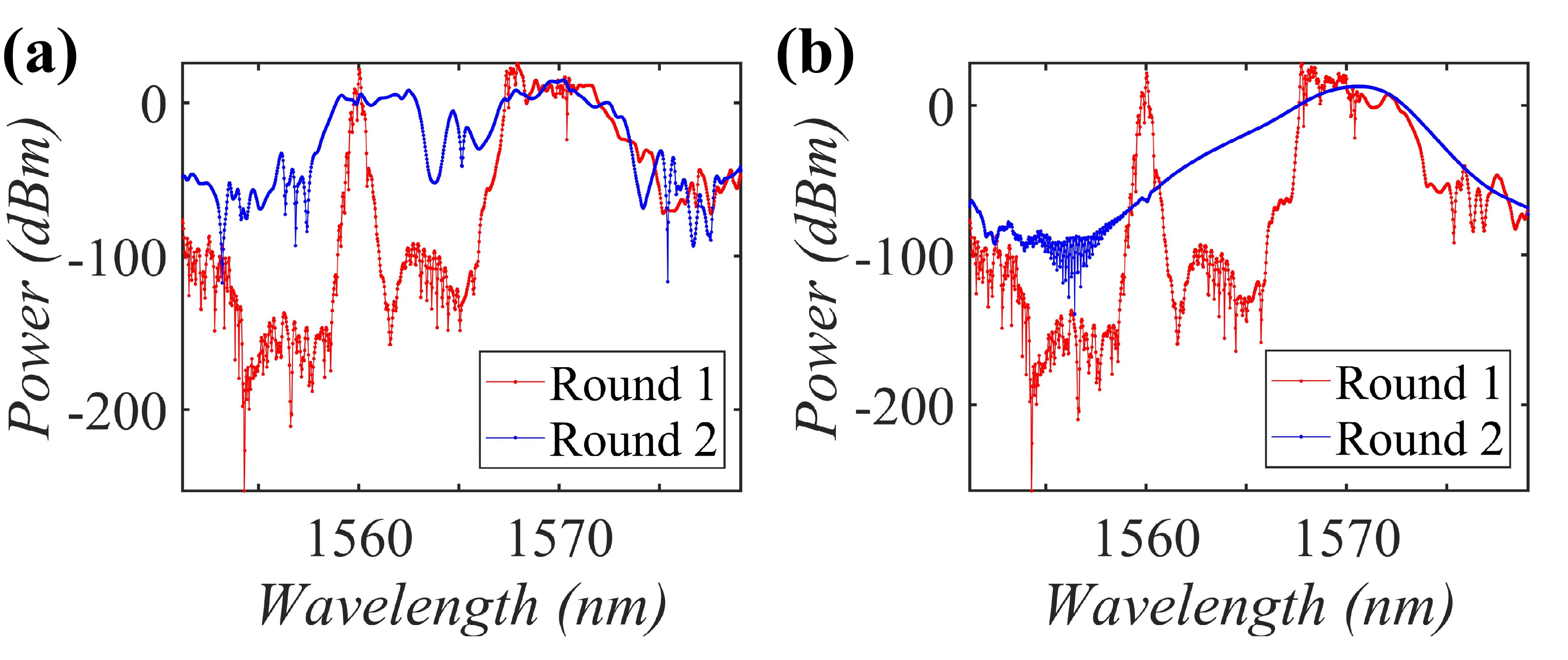}}
	\caption{Point-scan sampling optical spectrum of (a) Fig. 4(k) and (b) Fig. 4(h). }
\end{figure}

So far, the spectrum of the QML pulse has been preliminarily demonstrated to be a periodic oscillating state rather than a stabilized spectrum as demonstrated in most of the previous studies. To further clarify the reason for the emergence of the bound-state-like modulation, we simulate the point-scan sampling approach of the spectrometer to plot the spectrum in Fig. 4(k) and display the two rounds of scanning results in Fig. 6(a). The result, as obtained in the experiments, exhibits a bound-state-like modulation in a single scan, even though the spectrum itself may be smooth. The red line is the first scanning round. It contains part of the initiation process which is not visible in the actual optical spectrum because it evolves so fast. The blue line is the second scanning optical spectrum which successfully demonstrates the bound-state-like modulation in the numerical simulation. As a comparison, we also give the point-scan optical spectrum of Fig. 4(h) in Fig. 6(b), where the stabilized single wavelength can be clearly distinguished from the blue line.

Moreover, the clarity of this spectral oscillation phenomenon further supports our conjecture in reference \cite{22} that there exists a type of evolutionary path for mode-locked pulses from CWML to pulsating solitons to QML. We mean, if further deepen the modulation envelope strength of the pulsating soliton, the shape of the pulse keeps approaching QML.

In this study, we investigate experimentally and theoretically the dual-color QML pulse. Experimentally, we observed the dual-color QML pulses and found the bound-state-like envelope on their optical spectrum for the first time. Theoretically, a set of CGLEs based on hybrid mode-locked fiber lasers is established and numerically solved using the RK4 method to obtain the evolution of QML pulses. The simulation results match well with the experiment, proving that the QML pulse optical spectrum is periodic oscillatory, and two possible patterns of QML pulses are found. In addition, we have numerically simulated the bound-state-like modulation optical spectrum by simulating the point-scan sampling process of the spectrometer. Our results complete the gap of spectrum oscillatory behavior of QML for a long time, deepen the understanding of QML mechanism, and offer a foundation for the study of nonlinear evolutionary paths in fiber lasers.

\medskip

\noindent\textbf{Funding.} National Major Scientific Research Instrument Development Project of China (51927804); National Natural Science Foundation of China (62375220); Shaanxi Key Science and Technology Innovation Team Project (2023-CX-TD-06).

\medskip
\noindent\textbf{Disclosures.} The authors declare no conflicts of interest.

\medskip
\noindent\textbf{Data availability.} Data underlying the results presented in this paper are not publicly available at this time but may be obtained from the authors upon reasonable request. 

\bibliography{Ref}

\bibliographyfullrefs{Ref}

\end{document}